\def\rft#1{Table \ref{#1}}

\documentclass[prd,eqsecnum,leqnum,floats]{revtex4}

\begin{document}

\title{Spherical field in rotating space in 5D}

\author{ W. B. Belayev \\
 \normalsize Center for Relativity and Astrophysics,\\ \normalsize
185 Box, 194358, Sanct-Petersburg, Russia \\ \normalsize e-mail:
wbelayev@yandex.ru }

\begin{abstract}
A geodesic motion in rotating 5D space is studied in framework of
Kaluza-Klein theory. A proposed phenomenological model predicts
basic properties of the Pioneer-effect, namely, a) constant
additional acceleration of apparatus on distance from 20 to 50
a.e., b) its increase from 5 to 20 a.e., c) observed absence of
one in motion of planets.
\end{abstract}

\maketitle

\section{Introduction}

A five-dimensional model of the space-time was proposed by
Nordstrom \cite{Nordst} and Kaluza \cite{Kal} for unity of
gravitation and electromagnetism. Klein \cite{Klein} suggested a
compactification mechanism, owing to which internal space of the
Planck size forms additional dimension. In his theory a motion of
particle having rest mass in 4D can be described by equations of
null geodesic line in 5D, which are interpretation of massless
wave equation with some conditions.

In development of this model 5D space-time is considered as low
energy limit of more high-dimensional theories of supersymmetry,
supergravity and string theory. They admit scenario, in which
particle has a rest mass in 5D \cite{Overd, Leon1}. Exact
solutions of Kaluza-Klein and low limit of bosonic string theories
in 5D-6D \cite{Aguil} with toroidal compactification are
equivalent. Analogous conclusion is made in \cite{Leon2} with
comparison space-time-mass theory based on geometric properties of
5D space without compactification and braneworld model.
Predictions of five-dimensional model of extended space and its
experimental tests are considered in \cite{ Tsip1, Tsip2}.
Cosmological model with motion of matter in fifth dimension also
is examined \cite{Bel}. Astrophysical applications of braneworld
theories, including Arkani-Hamed-Dimopoulos-Dvali and
Einstein-Maxwell models with large extra dimensions, are analyzed
in \cite{Koyam}. EM model in 6D has become further development in
\cite{Kob}, where linear perturbations sourced by matter on the
brane are studied. In \cite{Arr} it is proposed low energy
effective theory on a regularized brane in 6D gauged chiral
supergravity. A possibility of rotations of particles in (4+n)D
space periodically returning in 4D is phenomenologically predicted
in ADD model \cite{ADD}, where radius of rotation defines the size
of additional dimension. Phenomena described by one-time physics
in 3+1 dimensions appear as various "shadows" similar phenomena
that occur in 4+2 dimensions with one extra space and one extra
time dimensions (more generally, d+2) \cite{BCQ, IBSC, Bars}.

In present paper it is considered some geometrical construction in
(4+1)D space-time with space-like fifth dimension and rotation in
4D spherical coordinates with transition thereupon to the standard
cylindric frame.  It is studied also (3+2)D space-time with
time-like additional dimension, where the motion is hyperbolic.

A motion of the particle in certain domain of space in appropriate
coordinates is assigned to be described with sufficient accuracy
by geodesic equations. Their solutions lead to conclusion that
rotation in 5D space-time exhibits itself in 4D as action of
central force. In Kaluza-Klein model this force divides into
components, the part of which associates with to electromagnetic
field.

In astrophysical applications proposed model of space-time is of
interest with respect to Pioneer effect. In some papers
\cite{Trenc,199}, see also review of efforts to explain anomaly
\cite{Tur}, presence of signal frequency bias is associated with
dependence of fundamental physical parameters from time. Though it
should allow for data \cite{And2}, which witness independence of
direction of additional acceleration from route of radial motion
with respect to Sun. A radial Rindler-like acceleration
\cite{dgrum,LIor} in itself also can't explain the Pioneer effect
because it is absent in planets motion \cite{Ior}.

\section{Geodesics in rotating space}

 Five-dimensional space-time having 4D spherical symmetry is considered in coordinate frame
 $X^{i}_s=(\tau,a,\theta,\varphi,\chi)$, where $a,\theta,\varphi,\chi$ are spherical space
 coordinates and $\tau=ct$, where $c$ is light velocity constant and $t$ is time. Rotating space-time with space-like fifth dimension \cite{Bel1} is described by metric
\begin{eqnarray}\label{f1}
 dS^2=[1-a^2B(a)^2]d{\tau}^2-da^2-a^2[2B(a)g(\chi)d\tau d\chi+\sin^2f(\chi)(d\theta^2+
 \sin^2\theta d\varphi^2)+ g^2(\chi)d\chi^2],
\end{eqnarray}
where function $f(\chi)$ is continuously increasing and it is
taken  $g=df/d\chi$. For the domain under review it is assumed
\begin{eqnarray}\label{f2}
B(a)=Ka^{-1/2},
\end{eqnarray}
where $K$ is constant.

Transition to five-dimensional cylindrical coordinates
$X^{i}_c=(\tau,r,\theta,\varphi,y)$ for $0\leq\chi\leq\pi$ is
performed by transformation
\begin{eqnarray}\label{f3}
r=a\sin f(\chi), \,\,\,\,\, y=a\cos f(\chi).
\end{eqnarray}

Geodesic equations in 5D for particle having rest mass are
\begin{eqnarray}\label{f332}
\frac{dU^{i}}{dS}+\Gamma^{i}_{kl}U^{k}U^{l}=0,
\end{eqnarray}
where $U^{i}$ are components of five-velocity vector and
$\Gamma^{i}_{kl}$ are 5D Christoffel symbols of second kind. In
spherical coordinates for metric (\ref{f1}) these equations, with
\begin{equation}\label{f31}
f(\chi)=\chi,
\end{equation}
take form
\begin{eqnarray}\label{f4}
\frac{d^{2}\tau}{dS^{2}}+\frac{K^2}{2}\frac{d\tau}{dS}\frac{da}{dS}
+\frac{Ka^{1/2}}{2}\frac{da}{dS}\frac{d\chi}{dS}
-\frac{Ka^{3/2}}{2}\sin(2\chi)\left(\frac{d\theta}{dS}\right)^{2}
-\frac{Ka^{3/2}}{2}\sin(2\chi)\sin^{2}\varphi\left(\frac{d\varphi}{dS}\right)^{2}=0,\\
\frac{d^{2}a}{dS^{2}}-\frac{K^2}{2}\left(\frac{d\tau}{dS}\right)^{2}
-\frac{3Ka^{1/2}}{2}\frac{d\tau}{dS}\frac{d\chi}{dS} -a
\sin^{2}\chi\left(\frac{d\theta}{dS}\right)^{2} -a \sin^{2}\theta
\sin^{2}\chi\left(\frac{d\varphi}{dS}\right)^{2}
-a\left(\frac{d\chi}{dS}\right)^{2}=0,\\
\frac{d^{2}\theta}{dS^{2}}+\frac{2}{a}\frac{da}{dS}\frac{d\theta}{dS}
+2 \cot\chi\frac{d\theta}{dS}\frac{d\chi}{dS} -\frac{\sin
(2\theta)}{2}\left(\frac{d\varphi}{dS}\right)^{2}=0,\\
\frac{d^{2}\varphi}{dS^{2}}+\frac{2}{a}\frac{da}{dS}\frac{d\varphi}{dS}
+2 \cot\theta\frac{d\theta}{dS}\frac{d\varphi}{dS} +2
\cot\chi\frac{d\varphi}{dS}\frac{d\chi}{dS}=0,\\
\frac{d^{2}\chi}{dS^{2}}+\frac{K(3-K^{2}a)}{2a^{3/2}}\frac{d\tau}{dS}\frac{da}{dS}
+\frac{4-K^{2}a}{2a}\frac{da}{dS}\frac{d\chi}{dS}
-\frac{(1-K^{2}a)\sin(2\chi)}{2}\left(\frac{d\theta}{dS}\right)^{2}-
\nonumber
\\ -\frac{(1-K^{2}a)\sin^{2}\theta
\sin(2\chi)}{2}\left(\frac{d\varphi}{dS}\right)^{2}=0.\label{f4a}
\end{eqnarray}
For particle with rest mass the solutions of these equations must
correspond to given by metric condition
\begin{equation}\label{f41}
1=(1-K^2a)U^{02}-2Ka^{3/2}U^0U^4-U^{12}-\sin^2\chi
a^2(U^{22}+\sin^{2}\theta U^{32})-a^2U^{42}.
\end{equation}

From the second equation of the system we obtain
\begin{equation}\label{f4121}
U^4=\frac{KU^0}{4a^{1/2}}\left\{-3+\mu\left[1+\frac{16}{aK^2U^{02}}\left(\frac{dU^1}{dS}-
a\sin^2\chi(U^{22}+\sin^2\theta
U^{32})\right)\right]^{1/2}\right\},
\end{equation}
where $\mu$ is $\pm 1$. We name corresponding solution a type I
for $\mu=-1$ and a type II for $\mu=1$.

When particles move along geodesics, which are arcs of circle:
\begin{equation}\label{f42}
U^1=U^2=U^3=0,
\end{equation}
equations of motion have following solutions:
\begin{equation}\label{f43}
U_{I}^0=\sigma, \,\,\,\,\, U_{I}^4=-\frac{\sigma K}{a^{1/2}}
\end{equation}
and
\begin{equation}\label{f51}
U_{II}^0=\frac{2\sigma}{\sqrt{4-K^{2}a}}, \,\,\,\,\,
U_{II}^4=-\frac{\sigma K}{a^{1/2}\sqrt{4-K^2a}},
\end{equation}
where $\sigma$ is $1,-1$.

Change of passage of time is defined as relation between intervals
of proper time $T=\int dS$ and coordinate time $\tau$. For
geodesic of type I for solution (\ref{f43}) chosen $\sigma=1$ we
have
\begin{equation}\label{f515}
d\tau=dT,
\end{equation}
i.e. time dilation is absence. With motion of particle along
circular geodesic of type II (\ref{f51}) we obtain
\begin{equation}\label{f6}
dT=\frac{1}{2}\sqrt{4-K^2 a}{d\tau}.
\end{equation}

\section{Representation in cylindric frame}

After substitutions of coordinate transformation being inverse to
(\ref{f3}), namely,
\begin{equation}\label{f7}
a=\sqrt{r^{2}+y^{2}},\,\,\,\,\,f(\chi)=\mathrm{arccot}\frac{y}{r}
\end{equation}
metric (\ref{f1}) with (\ref{f2}) is rewritten as
\begin{eqnarray}\label{f8}
dS^2=(1-K^2\sqrt{r^{2}+y^{2}})d{\tau}^2-2K(r^{2}
+y^{2})^{-1/4}d\tau(ydr-rdy)-dr^{2}-r^2(d\theta^2+ \sin^2\theta
d\varphi^2)-dy^2.
\end{eqnarray}
Geodesic equations will be
\begin{eqnarray}\label{f448}
\frac{d^{2}\tau}{dS^{2}}+\frac{{K^2} r}{2
\sqrt{{r^2}+{y^2}}}\frac{d\tau}{dS}\frac{dr}{dS}+ \frac{{K^2} y}{2
\sqrt{{r^2}+{y^2}}}\frac{d\tau}{dS}\frac{dy}{dS} +\frac{K r y}{2
({r^2}+{y^2})^{5/4}}\left(\frac{dr}{dS}\right)^{2}
-\frac{K({r^2}-{y^2})}{2
({r^2}+{y^2})^{5/4}}\frac{dr}{dS}\frac{dy}{dS} -\nonumber\\
-\frac{K r
y}{({r^2}+{y^2})^{1/4}}\left(\frac{d\theta}{dS}\right)^{2}
-\frac{K r y
}{({r^2}+{y^2})^{1/4}}\sin^{2}\varphi\left(\frac{d\varphi}{dS}\right)^{2}
-\frac{K r y}{2({r^2}+{y^2})^{5/4}}\left(\frac{dy}{dS}\right)^{2}
=0,\\ \frac{d^{2}r}{dS^{2}}-\frac{{K^2} r}{2
\sqrt{{r^2}+{y^2}}}\left(\frac{d\tau}{dS}\right)^{2} -\frac{ {K^3}
r y}{2 ({r^2}+{y^2})^{3/4}}\frac{d\tau}{dS}\frac{dr}{dS}
-\frac{{K^3}{y^2}-3K
\sqrt{{r^2}+{y^2}}}{2({r^2}+{y^2})^{3/4}}\frac{d\tau}{dS}\frac{dy}{dS}-\nonumber\\
-\frac{{K^2} r {y^2}}{2
({r^2}+{y^2})^{3/2}}\left(\frac{dr}{dS}\right)^{2}
 +\frac{K^2 y
({r^2}-{y^2})}{2 ({r^2}+{y^2})^{3/2}}\frac{dr}{dS}\frac{dy}{dS} +
\frac{r(K^2
y^2-\sqrt{r^2+y^2})}{\sqrt{{r^2}+{y^2}}}\left(\frac{d\theta}{dS}\right)^{2}+\nonumber\\
+ \frac{r({K^2}
{y^2}-\sqrt{r^2+y^2})}{\sqrt{{r^2}+{y^2}}}\sin^{2}\theta\left(\frac{d\varphi}{dS}\right)^{2}
+ \frac{{K^2} r {y^2}}{2
({r^2}+{y^2})^{3/2}}\left(\frac{dy}{dS}\right)^{2}=0,\label{f448b}\\
\frac{d^{2}\theta}{dS^{2}}+\frac{2}{r}\frac{dr}{dS}\frac{d\theta}{dS}
-\frac{\sin (2\theta)}{2}\left(\frac{d\varphi}{dS}\right)^{2}=0,\\
\frac{d^{2}\varphi}{dS^{2}}+\frac{2}{r}\frac{dr}{dS}\frac{d\varphi}{dS}
+2 \cot\theta\frac{d\theta}{dS}\frac{d\varphi}{dS}=0,\\
\frac{d^{2}y}{dS^{2}}-\frac{{K^2} y}{2
\sqrt{{r^2}+{y^2}}}\left(\frac{d\tau}{dS}\right)^{2} +\frac{{K^3}
{r^2}-3K
\sqrt{{r^2}+{y^2}}}{2({r^2}+{y^2})^{3/4}}\frac{d\tau}{dS}\frac{dr}{dS}+\frac{{K^3}
r y}{2 ({r^2}+{y^2})^{3/4}}\frac{d\tau}{dS}\frac{dy}{dS}+\nonumber
\\ +\frac{{K^2} y {r^2}}{2
({r^2}+{y^2})^{3/2}}\left(\frac{dr}{dS}\right)^{2}-\frac{K^2 r
({r^2}-{y^2})}{2 ({r^2}+{y^2})^{3/2}}\frac{dr}{dS}\frac{dy}{dS} -
\frac{K^2 r^2
y}{\sqrt{{r^2}+{y^2}}}\left(\frac{d\theta}{dS}\right)^{2} -
\frac{K^2 r^2
y}{\sqrt{{r^2}+{y^2}}}\sin^{2}\theta\left(\frac{d\varphi}{dS}\right)^{2}
-\nonumber \\ - \frac{K^2 r^2 y}{2
({r^2}+{y^2})^{3/2}}\left(\frac{dy}{dS}\right)^{2}=0.\label{f448a}
\end{eqnarray}

Components of five-velocity vector corresponding to coordinates
$r$ and $y$ are found by differentiation of transformation
(\ref{f3}) and will be
\begin{equation}\label{f9}
V^1=\sin\chi U^1+a\cos\chi U^4,\,\,\,\,\, V^4=\cos\chi U^1-a
\sin\chi U^4 .
\end{equation}
Condition given by metric (\ref{f8}) for the time-like path is
\begin{equation}\label{f418}
1=(1-K^2\sqrt{r^{2}+y^{2}})V^{02}-2K(r^{2}
+y^{2})^{-1/4}V^{0}(yV^1-rV^4)-V^{12}-r^2(V^{22}+ \sin^2\theta
V^{32})-V^{42}.
\end{equation}

In coordinate frame $X_c$ non-zero components of five-velocity
vector corresponding to solutions of geodesic equations in
coordinates $X_s$ (\ref{f42})-(\ref{f51}) of types I and II is
rewritten as
\begin{eqnarray}\label{f10}
V_{I}^0=\sigma, \,\,\,\,\, V_{I}^1=-\frac{\sigma
Ky}{(r^2+y^2)^{1/4}},\,\,\,\,\,V_{I}^4=\frac{\sigma
Kr}{(r^2+y^2)^{1/4}},\\
V_{II}^0=\frac{2\sigma}{(4-K^2\sqrt{r^2+y^2})^{1/2}}, \\
V_{II}^1=-\frac{\sigma
Ky}{(r^2+y^2)^{1/4}(4-K^2\sqrt{r^2+y^2})^{1/2}},\\
V_{II}^4=\frac{\sigma
Kr}{(r^2+y^2)^{1/4}(4-K^2\sqrt{r^2+y^2})^{1/2}}.\label{f10a}
\end{eqnarray}
For $y=0$ $(\chi=\pi/2)$ they correspond with stationary in 4D
particle.

For motion in the neighborhood of point $(\tau_0, r_0,\pi/2,0,0)$
with conditions $V^2_0=V^3_0=0$ geodesic equations are reduced to
\begin{eqnarray}\label{f112}
\frac{d^{2}\tau}{dS^{2}}+\frac{{K^2}}{2}\frac{d\tau}{dS}\frac{dr}{dS}-\frac{K}{2r_0^{1/2}}\frac{dr}{dS}\frac{dy}{dS}=0,\\
\frac{d^{2}r}{dS^{2}}-\frac{{K^2}}{2}\left(\frac{d\tau}{dS}\right)^{2}
+\frac{3K}{2r_0^{1/2}}\frac{d\tau}{dS}\frac{dy}{dS}=0,\\
\frac{d^{2}y}{dS^{2}}+\frac{{K^3}r_0-3K}{2r_0^{1/2}}\frac{d\tau}{dS}\frac{dr}{dS}-\frac{K^2}{2}\frac{dr}{dS}\frac{dy}{dS}=0.\label{f112a}
\end{eqnarray}
Condition (\ref{f418}) takes form
\begin{equation}\label{f4188}
1=(1-K^2r_0)V^{02}-2Kr_0^{1/2}V^{0}V^4-V^{12}-V^{42}.
\end{equation}
For circular motion (\ref{f10})-(\ref{f10a}) Eq. (\ref{f448b})
yields radial accelerations
\begin{eqnarray}\label{f64}
\frac{dV_{I}^1}{dS}=-K^2,\\
\frac{dV_{II}^1}{dS}=-\frac{K^2}{4-K^2\sqrt{r^2+y^2}}.
\end{eqnarray}

\section{Metrics with time-like fifth coordinate}

A space-time having hyperbolic motion with coordinates
$\check{X}^{i}_g=(\check{\tau},\check{a},\check{\theta},\check{\varphi},\check{\chi})$
is described by metric
\begin{eqnarray}\label{f12}
 dS^2=(1+\check{K}^2\check{a})d{\check{\tau}}^2-d\check{a}^{2}+
 \check{a}^2[2\check{K}\check{a}^{-1/2}d\check{\tau} d\check{\chi}-\cosh \check{\chi}^2(d\check{\theta}^2+
 \sin^2\check{\theta} d\check{\varphi}^2)+ d\check{\chi}^2],
\end{eqnarray}
where $\check{\chi}$ is assumed to be time-like and $\check{K}$ is
constant. This metric can be obtained from (\ref{f1}) for
(\ref{f2}), (\ref{f31}) by substitution $K=-{\rm i}\check{K},\,\,
\chi=\frac{\pi}{2}-{\rm i}\check{\chi}$ and addition of
$(\check{})$ in the notation of other coordinates.

The geodesics equations for a particle motion along time-like path
are
\begin{eqnarray}\label{f449}
\frac{d^{2}\check{\tau}}{dS^{2}}-\frac{\check{K}^2}{2}\frac{d\check{\tau}}{dS}\frac{d\check{a}}{dS}
-\frac{\check{K}\check{a}^{1/2}}{2}\frac{d\check{a}}{dS}\frac{d\check{\chi}}{dS}
-\frac{\check{K}\check{a}^{3/2}}{2}\sinh(2\check{\chi})\left(\frac{d\check{\theta}}{dS}\right)^{2}
-\frac{\check{K}\check{a}^{3/2}}{2}\sinh(2\check{\chi})\sin^{2}\check{\varphi}\left(\frac{d\check{\varphi}}{dS}\right)^{2}=0,\\
\frac{d^{2}\check{a}}{dS^{2}}+\frac{\check{K}^2}{2}\left(\frac{d\check{\tau}}{dS}\right)^{2}
+\frac{3\check{K}\check{a}^{1/2}}{2}\frac{d\check{\tau}}{dS}\frac{d\check{\chi}}{dS}
-\check{a}
\cosh^{2}\check{\chi}\left(\frac{d\check{\theta}}{dS}\right)^{2}
-\check{a} \sin^{2}\check{\theta}
\cosh^{2}\check{\chi}\left(\frac{d\check{\varphi}}{dS}\right)^{2}
+\check{a}\left(\frac{d\check{\chi}}{dS}\right)^{2}=0,\\
\frac{d^{2}\check{\theta}}{dS^{2}}+\frac{2}{\check{a}}\frac{d\check{a}}{dS}\frac{d\check{\theta}}{dS}
+2
\tanh\check{\chi}\frac{d\check{\theta}}{dS}\frac{d\check{\chi}}{dS}
-\frac{\sin
(2\check{\theta})}{2}\left(\frac{d\check{\varphi}}{dS}\right)^{2}=0,\\
\frac{d^{2}\check{\varphi}}{dS^{2}}+\frac{2}{\check{a}}\frac{d\check{a}}{dS}\frac{d\check{\varphi}}{dS}
+2
\cot\check{\theta}\frac{d\check{\theta}}{dS}\frac{d\check{\chi}}{dS}
+2
\tanh\check{\chi}\frac{d\check{\varphi}}{dS}\frac{d\check{\chi}}{dS}=0,\\
\frac{d^{2}\check{\chi}}{dS^{2}}+\frac{\check{K}(3+\check{K}^{2}\check{a})}{2\check{a}^{3/2}}\frac{d\check{\tau}}{dS}\frac{d\check{a}}{dS}
+\frac{4+\check{K}^{2}\check{a}}{2\check{a}}\frac{d\check{a}}{dS}\frac{d\check{\chi}}{dS}
+\frac{(1+\check{K}^{2}\check{a})\sinh(2\check{\chi})}{2}\left(\frac{d\check{\theta}}{dS}\right)^{2}+\nonumber\\
+\frac{(1+\check{K}^{2}\check{a})\sin^{2}\check{\theta}
\sinh(2\check{\chi})}{2}\left(\frac{d\check{\varphi}}{dS}\right)^{2}=0.\label{f449a}
\end{eqnarray}
For particle with rest mass the solutions of these equations must
correspond to given by metric condition
\begin{equation}\label{f414}
1=(1+\check{K}^2\check{a})\check{U}^{02}+2\check{K}\check{a}^{3/2}\check{U}^0\check{U}^4-\check{U}^{12}-\cosh^2\check{\chi}
\check{a}^2(\check{U}^{22}+\sin^{2}\check{\theta}
\check{U}^{32})+\check{a}^2\check{U}^{42}.
\end{equation}
Second equation of the system yields
\begin{equation}\label{f41212}
\check{U}^4=\frac{\check{K}\check{U}^0}{4\check{a}^{1/2}}\left\{-3
+\mu\left[1-\frac{16}{\check{a}\check{K}^2\check{U}^{02}}\left(\frac{d\check{U}^1}{dS}
-\check{a}\cosh^2\check{\chi}(\check{U}^{22}+\sin^2\check{\theta}\check{U}^{32})\right)\right]^{1/2}\right\}.
\end{equation}

With hyperbolic motion for $\check{U}^1=\check{U}^2=\check{U}^3=0$
corresponding five-velocity vectors have non-zero components
\begin{eqnarray}\label{f433}
\check{U}_{I}^0=\sigma, \,\,\,\,\, \check{U}_{I}^4=-\frac{\sigma
\check{K}}{\check{a}^{1/2}},\\
\check{U}_{II}^0=\frac{2\sigma}{\sqrt{4+\check{K}^{2}\check{a}}},
\,\,\,\,\, \check{U}_{II}^4=-\frac{\sigma
\check{K}}{\check{a}^{1/2}\sqrt{4+\check{K}^2\check{a}}}.\label{f4334}
\end{eqnarray}
For solution of type I time dilation is absent:
\begin{equation}\label{f443}
d\check{T}=d\check{\tau},
\end{equation}
and for type II increase of proper time passage is given by
\begin{equation}\label{f44}
d\check{T}=\frac{1}{2}\sqrt{4+\check{K}^2 \check{a}}d\check{\tau}.
\end{equation}

Transition to cylindrical coordinates
$\check{X}^{i}_{c}=(\check{\tau},\check{r},\check{\theta},\check{\varphi},\check{y})$
is realized by transformation
\begin{equation}\label{f45}
\check{r}=\check{a}\cosh\check{\chi},\,\,\,\,\,\check{y}=\check{a}\sinh\check{\chi}.
\end{equation}
Corresponding components of five-velocity vector are
\begin{equation}\label{f46}
\check{V}^1=\cosh\check{\chi}
\check{U}^1+\check{a}\sinh\check{\chi}\check{U}^4,\,\,\,\,\,
\check{V}^4=\sinh\check{\chi} \check{U}^1+\check{a}
\cosh\check{\chi} \check{U}^4.
\end{equation}
Inverse coordinate transformation is written as
\begin{equation}\label{f47}
\check{a}=\sqrt{\check{r}^{2}-\check{y}^{2}},\,\,\,\,\,
\check{\chi}=\mathrm{arcoth} \frac{\check{y}}{\check{r}}.
\end{equation}
Substituting this in (\ref{f12}) gives
\begin{eqnarray}\label{f48}
dS^2=(1+\check{K}^2\sqrt{\check{r}^{2}-\check{y}^{2}})d{\check{\tau}}^2-
2\check{K}(\check{r}^{2}-\check{y}^{2})^{-1/4}d\check{\tau}(\check{y}d\check{r}-\check{r}d\check{y})
-d\check{r}^{2}-\check{r}^2(d\check{\theta}^2+
\sin^2\check{\theta} d\check{\varphi}^2)+d\check{y}^2.
\end{eqnarray}
The same line element can be obtained by replacement $K=-{\rm
i}\check{K},\,\, y={\rm i}\check{y}$ and addition of $(\check{})$
in the notation of other coordinates in (\ref{f8}).

Geodesics equations for motion of particle having rest mass is
written as
\begin{eqnarray}\label{f500}
\frac{d^{2}\check{\tau}}{dS^{2}}-\frac{{\check{K}^2} \check{r}}{2
\sqrt{{\check{r}^2}-{\check{y}^2}}}\frac{d\check{\tau}}{dS}\frac{d\check{r}}{dS}+
\frac{{\check{K}^2} \check{y}}{2
\sqrt{{\check{r}^2}-{\check{y}^2}}}\frac{d\check{\tau}}{dS}\frac{d\check{y}}{dS}
+\frac{\check{K} \check{r}
\check{\check{y}}}{2({\check{r}^2}-{\check{y}^2})^{5/4}}\left(\frac{d\check{r}}{dS}\right)^{2}
-\frac{\check{K} ({\check{r}^2}+{\check{y}^2})}{2({\check{r}^2}-
{\check{y}^2})^{5/4}}\frac{d\check{r}}{dS}\frac{d\check{y}}{dS}-\nonumber
\\ -\frac{\check{K} \check{r} \check{y}}{c
({\check{r}^2}-{\check{y}^2})^{1/4}}\left(\frac{d\check{\theta}}{dS}\right)^{2}
-\frac{\check{K} \check{r} \check{y} }{
({\check{r}^2}-{\check{y}^2})^{1/4}}\sin^{2}\check{\varphi}\left(\frac{d\check{\varphi}}{dS}\right)^{2}
+\frac{\check{K} \check{r} \check{y}}{2
({\check{r}^2}-{\check{y}^2})^{5/4}}\left(\frac{d\check{y}}{dS}\right)^{2}
=0,
\\ \frac{d^{2}\check{r}}{dS^{2}}+\frac{{\check{K}^2}
\check{r}}{2
\sqrt{{\check{r}^2}-{\check{y}^2}}}\left(\frac{d\check{\tau}}{dS}\right)^{2}
+\frac{ {\check{K}^3} \check{r} \check{y}}{2
({\check{r}^2}-{\check{y}^2})^{3/4}}\frac{d\check{\tau}}{dS}\frac{d\check{r}}{dS}
-\frac{\check{K}^3\check{y}^2-3\check{K}\sqrt{\check{r}^2-\check{y}^2}}{2
({\check{r}^2}-{\check{y}^2})^{3/4}}\frac{d\check{\tau}}{dS}\frac{d\check{y}}{dS}
-\nonumber\\ -\frac{{\check{K}^2} \check{r} {\check{y}^2}}{2
({\check{r}^2}-{\check{y}^2})^{3/2}}\left(\frac{d\check{r}}{dS}\right)^{2}
+\frac{\check{K}^2 \check{y} ({\check{r}^2}+{\check{y}^2})}{2
({\check{r}^2}-{\check{y}^2})^{3/2}}\frac{d\check{r}}{dS}\frac{d\check{y}}{dS}
+ \frac{\check{r}({\check{K}^2}
{\check{y}^2}-\sqrt{{\check{r}^2}-{\check{y}^2}})}{\sqrt{{\check{r}^2}-{\check{y}^2}}}\left(\frac{d\check{\theta}}{dS}\right)^{2}
+\nonumber\\+ \frac{\check{r}(\check{K}^2
\check{y}^2-\sqrt{\check{r}^2-\check{y}^2})}{\sqrt{{\check{r}^2}
-{\check{y}^2}}}\sin^{2}\check{\theta}\left(\frac{d\check{\varphi}}{dS}\right)^{2}
-\frac{{\check{K}^2} \check{r} {\check{y}^2}}{2
({\check{r}^2}-{\check{y}^2})^{3/2}}\left(\frac{d\check{y}}{dS}\right)^{2}=0,\label{f500a}\\
\frac{d^{2}\check{\theta}}{dS^{2}}+\frac{2}{\check{r}}\frac{d\check{r}}{dS}\frac{d\check{\theta}}{dS}
-\frac{\sin
(2\check{\theta})}{2}\left(\frac{d\check{\varphi}}{dS}\right)^{2}=0,\\
\frac{d^{2}\check{\varphi}}{dS^{2}}+\frac{2}{\check{r}}\frac{d\check{r}}{dS}\frac{d\check{\varphi}}{dS}
+2
\cot\check{\theta}\frac{d\check{\theta}}{dS}\frac{d\check{\varphi}}{dS}=0,\\
\frac{d^{2}\check{y}}{dS^{2}}+\frac{\check{K^2}\check{y}}{2
\sqrt{{\check{r}^2}-{\check{y}^2}}}\left(\frac{d\check{\tau}}{dS}\right)^{2}
+\frac{{\check{K}^3} {\check{r}^2}+3\check{K}
\sqrt{{\check{r}^2}-{\check{y}^2}}}{2
({\check{r}^2}-{\check{y}^2})^{3/4}}\frac{d\check{\tau}}{dS}\frac{d\check{r}}{dS}
-\frac{ {\check{K}^3}\check{r} \check{y}}{2
({\check{r}^2}-{\check{y}^2})^{3/4}}\frac{d\check{\tau}}{dS}\frac{d\check{y}}{dS}-\nonumber
\\ - \frac{{\check{K}^2} \check{y} {\check{r}^2}}{2 ({\check{r}^2}-
{\check{y}^2})^{3/2}}\left(\frac{d\check{r}}{dS}\right)^{2}
+\frac{\check{K}^2 \check{r} ({\check{r}^2}+{\check{y}^2})}{2
({\check{r}^2}-{\check{y}^2})^{3/2}}\frac{d\check{r}}{dS}\frac{d\check{y}}{dS}
+ \frac{\check{K}^2 \check{r}^2
\check{y}}{\sqrt{{\check{r}^2}-{\check{y}^2}}}\left(\frac{d\check{\theta}}{dS}\right)^{2}
+ \frac{\check{K}^2 \check{r}^2
\check{y}}{\sqrt{{\check{r}^2}-{\check{y}^2}}}\sin^{2}\check{\theta}\left(\frac{d\check{\varphi}}{dS}\right)^{2}
-\nonumber \\ - \frac{\check{K}^2 \check{r}^2 \check{y}}{2
({\check{r}^2}-
{\check{y}^2})^{3/2}}\left(\frac{d\check{y}}{dS}\right)^{2}=0.\label{f5001}
\end{eqnarray}
Condition given by metric (\ref{f48}) for the time-like path is
\begin{equation}\label{f481}
1=(1+\check{K}^2\sqrt{\check{r}^{2}-\check{y}^{2}})\check{V}^{02}-2\check{K}(\check{r}^{2}
-\check{y}^{2})^{-1/4}\check{V}^{0}(\check{y}\check{V}^1-\check{r}\check{V}^4)-\check{V}^{12}-\check{r}^2(\check{V}^{22}+
\sin^2\check{\theta} \check{V}^{32})+\check{V}^{42}.
\end{equation}

A non-zero components of five-velocity vectors corresponding
hyperbolic solutions (\ref{f433}), (\ref{f4334}) are
\begin{eqnarray}\label{f49}
\check{V}_{I}^0=\sigma, \,\,\,\,\, \check{V}_{I}^1=-\frac{\sigma
\check{K}\check{y}}{(\check{r}^2
-\check{y}^2)^{1/4}},\,\,\,\,\,\check{V}_{I}^4=-\frac{\sigma
\check{K}\check{r}}{(\check{r}^2-\check{y}^2)^{1/4}},\\
\check{V}_{II}^0=\frac{2\sigma}{(4+\check{K}^2\sqrt{\check{r}^2-\check{y}^2})^{1/2}},
\\ \check{V}_{II}^1=-\frac{\sigma
\check{K}\check{y}}{(\check{r}^2-\check{y}^2)^{1/4}(4+
\check{K}^2\sqrt{\check{r}^2-\check{y}^2})^{1/2}},\\
\check{V}_{II}^4=-\frac{\sigma
\check{K}\check{r}}{(\check{r}^2-\check{y}^2)^{1/4}(4+\check{K}^2\sqrt{\check{r}^2-
\check{y}^2})^{1/2}}.\label{f499}
\end{eqnarray}

For motion in the neighborhood of point $(\check{\tau}_0,
\check{r}_0,\pi/2,0,0)$ with conditions
$\check{V}^2_0=\check{V}^3_0=0$ local solution is found from
reduced Eqs. (\ref{f500})-(\ref{f5001}), which turned out to
\begin{eqnarray}\label{f492}
\frac{d^{2}\check{\tau}}{dS^{2}}-\frac{{\check{K}^2}}{2}\frac{d\check{\tau}}{dS}\frac{d\check{r}}{dS}
-\frac{\check{K}}{2\check{r}_0^{1/2}}\frac{d\check{r}}{dS}\frac{d\check{y}}{dS}=0,\\
\frac{d^{2}\check{r}}{dS^{2}}+\frac{{\check{K}^2}}{2}\left(\frac{d\check{\tau}}{dS}\right)^{2}+
\frac{3\check{K}}{2\check{r}_0^{1/2}}\frac{d\check{\tau}}{dS}\frac{d\check{y}}{dS}=0,\label{f492b}\\
\frac{d^{2}\check{y}}{dS^{2}}
+\frac{\check{K}^3\check{r}_0+3\check{K}}{2\check{r}_0^{1/2}}\frac{d\check{\tau}}{dS}\frac{d\check{r}}{dS}
+\frac{\check{K}^2}{2}\frac{d\check{r}}{dS}\frac{d\check{y}}{dS}=0.\label{f492a}
\end{eqnarray}
Condition (\ref{f481}) takes form
\begin{equation}\label{f428}
1=(1+\check{K}^2
\check{r}_0)\check{V}^{02}+2\check{K}\check{r}_0^{1/2}\check{V}^0\check{V}^4-\check{V}^{12}
+\check{V}^{42}.
\end{equation}
For hyperbolic motion Eqs. (\ref{f49})-(\ref{f499}) radial
accelerations are
\begin{eqnarray}\label{f652}
\frac{d\check{V}_{I}^1}{dS}=\check{K}^2,
\end{eqnarray}
\begin{eqnarray}\label{f651}
\frac{d\check{V}_{II}^1}{dS}=\frac{\check{K}^2}{4+\check{K}^2\sqrt{\check{r}^2-\check{y}^2}}.
\end{eqnarray}

\section{Kaluza-Klein model}

In Kaluza-Klein theory the line element is brought in form
\begin{equation}\label{f512}
d^{2}S=g_{ij}dx^{i}dx^{j}+\varepsilon
\Phi^{2}(A_{i}dx^{i}+\varepsilon dy)^{2},
\end{equation}
where $x^{i}$ and $g_{ij}$ is coordinates and metrical tensor of
4D space-time, $\Phi$ and $A$ are scalar and vector potential.
Metrical coefficients and potentials are functions of $x^{i}$ and
$y$. Constant $\varepsilon$ equals 1 for time-like fifth
coordinate $y$ and -1, when it is space-like.

In this form metrics (\ref{f8}) and (\ref{f48}) are represented by
line-element of 4D space-time
\begin{equation}\label{f52}
ds^{2}=\left(1-K^{2}\frac{y^{2}}{a_{\varepsilon}}\right)d\tau^{2}-2Ka_{\varepsilon}^{-1/2}y
d\tau dr-dr^{2}-r^{2}(d\theta^{2}+\sin^{2}\theta d\varphi^{2})
\end{equation}
and potentials
\begin{equation}\label{f53}
A_0=Kra^{-1/2}_{\varepsilon},\,\,\,\,\,A_1=A_2=A_3=0
,\,\,\,\,\,\Phi=1,
\end{equation}
where it is denoted $a_{\varepsilon}=\sqrt{r^{2}-\varepsilon
y^{2}}$.

If 4D metric satisfies cylindrical conditions $\frac{\partial
g_{ij}}{\partial y}=0$ electromagnetic field is defined by
$F_{ij}=\partial_{i}A_{j}-\partial_{j}A_{i}.$ Ratio of electric
charge to mass in 4D is written as
\begin{equation}\label{54}
\frac{q}{m}=\frac{Q}{\sqrt{1-Q^{2}}},
\end{equation}
with scalar function
\begin{equation}\label{f55}
Q=\varepsilon
\Phi^{2}\left(\frac{dy}{dS}+A_{i}\frac{dx^{i}}{dS}\right).
\end{equation}
In more general case with 4D metrical coefficients being dependent
on $y$ the relationship between $Q$ and $q/m$ is not identical
\cite{Overd}, but value $Q=0$ also corresponds to $q=0$.

For considering space-time after substituting components of
five-velocity vector (\ref{f10}) and (\ref{f49}) we obtain for the
geodesics of the type I:
\begin{equation}\label{f56}
Q_I=0.
\end{equation}
This value is interpreted as neutral charge of a test particle.
For the type II scalar function is
\begin{equation}\label{f57}
Q_{II}=\frac{\varepsilon\sigma
Kr}{a_{\varepsilon}^{1/2}\sqrt{4+\varepsilon
K^{2}a_{\varepsilon}}}.
\end{equation}

The light trajectory is assumed to be isotropic curve both in 4D
and in 5D: ds=dS=0. From (\ref{f512})-(\ref{f53}) we obtain
solutions
\begin{equation}\label{f58}
\frac{dy}{d\tau}=0,
\end{equation}
and
\begin{equation}\label{f59}
\frac{dy}{d\tau}=-\varepsilon\frac{2Kr}{a_{\varepsilon}^{1/2}}.
\end{equation}

\section{Astrophysical applications}

Considering phenomenology of particles motion in 5D we assume that
stationary in 3D space particles, having rest mass, move in
spherical or hyperbolic frames in 5D along geodesics with constant
radial coordinate: (\ref{f43}), (\ref{f51}) and
(\ref{f10})-(\ref{f10a}) or (\ref{f433}), (\ref{f4334}) and
(\ref{f49})-(\ref{f499}). It is suggested also that in cylindrical
frame matter moves along fifth coordinate in single direction,
which is opposite to antimatter motion.

In case of space-like fifth dimension function $f$ from metric
(\ref{f1}) is chosen so that its meaning is continuously
increasing in intervals $I_n^{+}=[2\pi n,\,\,\pi+2\pi n]$ for integer
$n$. Since value $r<0$ is inadmissible in cylindric frame we must
assume that function $f$ has discontinuity on the endpoints of
$I_n$, which prescribes singularity. It can be avoid if model of
binary world consisting of universe - anti-universe pair
\cite{Lind, Dub, Rand} is considered under the assumption that it
possesses a large number of copies \cite{Ark,DSV}, in which a
physical laws are identical. In bulk  a
space-time half $I_n^{-}=[-\pi+2\pi n,\,\, 2\pi n]$ put into
accordance with packet of 4D anti-universes. With condition (\ref{f31}) intervals
$I_n^{+},\,\,\,\,I_n^{-}$ contain values of additional coordinate $\chi$. Rotation of one
particle with transition to cylindrical coordinates should be
interpreted as motion of particle and anti-particle through
opposite packets of branes, which conforms to CPT-symmetry of the
universe and anti-universe. Thus a birth of the pair
particle-antiparticle is assumed to occur in points $y=-+a_0,\,\,
r=0$, after which they move through opposite packets of branes and
annihilate, when $y=+-a_1,\,\, r=0$.

\subsection{Basic properties of Pioneer effect model}

Recently much attention was attracted to the Pioneer effect, which
consists in additional acceleration of spacecrafts Pioneer 10/11
\cite{Tur,And2,And1,And3} $a_p=(8.74\pm 1.33)\times 10^{-8}$ cm
s$^{-2}$ directed to the inner part of the solar system. We will
analyze how much studying models of rotating space conform to this
data. Motion of the spacecrafts and the planets is considered in
the frame of the Sun \cite{Tur1}.

For this analysis we must use geodesics of the first type because,
as was shown in Sec. 5, for geodesics with constant radial
coordinate they correspond to the neutrally charged particles.
Their proper time coincides with coordinate time for trajectories,
which are the arcs of circle (\ref{f515}) or hyperbola
(\ref{f443}). Also motion of light is assumed to correspond with
equation (\ref{f58}), i.e. a light shift along the fifth
coordinate is absent.

In the Sun's gravity field motion of the particle with rest mass
is described approximately by equations
\begin{equation}\label{f61}
\frac{dV^{i}}{dS}+\Gamma^{i}_{kl}V^{k}V^{l}=G^i,
\end{equation}
where $G^i$ is gravity force vector without part related to space
rotation and left terms correspond to Eqs.
(\ref{f448})-(\ref{f448a}) or (\ref{f500})-(\ref{f5001}). Denoting acceleration
$W^i=\frac{dV^{i}}{dS}$ we divide it into $W^i=W^i_g+W^i_F$, where
$W^i_F$ conforms in case $G^2=0$, $\theta=\pi/2$ in the
neighborhood of point $(\tau_0, r_0,\pi/2,0,0)$ to equations
\begin{eqnarray}\label{f62}
W_F^0=G^0,\\
W_{F}^1=G^1+r_0\left(\frac{d\varphi}{dS}\right)^{2},\\ W_F^2=0,\\
W_{F}^3=G^3-\frac{2}{r_0}\frac{dr}{dS}\frac{d\varphi}{dS},\\
W_{F}^4=G^4.
\end{eqnarray}
Accelerations $W^i_g$ correspond to Eqs.
(\ref{f112})-(\ref{f112a}) or (\ref{f492})-(\ref{f492a}).

By using analogy with motion of particle in central gravity field
in 4D, we take $U_I^1$ and $\,dU_I^1/dS\,$ for $\,\chi=\pi/2\,$ in
spherical coordinates or $\,\check{U}_I^1\,$ and
$\,d\check{U}_I^1/dS\,$ for $\check{\chi}=0$ in hyperbolic
coordinates as radial velocity and acceleration observed in 4D
surface of five-dimensional space-time.

\subsection{Model with space-like fifth coordinate}

In spherical coordinates in the neighborhood of point $(\tau_0,
r_0,\pi/2,0,\pi/2)$ Eqs. (\ref{f112})-(\ref{f112a}) correspond to
system (\ref{f4})-(\ref{f4a}) reduced to
\begin{eqnarray}\label{f633}
\frac{d^{2}\tau}{dS^{2}}+\frac{K^2}{2}\frac{d\tau}{dS}\frac{da}{dS}
+\frac{Kr_0^{1/2}}{2}\frac{da}{dS}\frac{d\chi}{dS}=0,\\
\frac{d^{2}a}{dS^{2}}-\frac{K^2}{2}\left(\frac{d\tau}{dS}\right)^{2}
-\frac{3r_0^{1/2}}{2}\frac{d\tau}{dS}\frac{d\chi}{dS}
-r_0\left(\frac{d\chi}{dS}\right)^{2}=0,\\
\frac{d^{2}\chi}{dS^{2}}+\frac{K(3-K^{2}r_0)}{2r_0^{3/2}}\frac{d\tau}{dS}\frac{da}{dS}
+\frac{4-K^{2}r_0}{2r_0}\frac{da}{dS}\frac{d\chi}{dS}=0.\label{f6334}
\end{eqnarray}
For closed to circular motion (\ref{f42}), (\ref{f43})
non-vanishing five-velocities are written in form
\begin{eqnarray}\label{f63}
U_{I}^0=\sigma+\alpha^{0},\\ U_{I}^1=\alpha^{1},\\
U_{I}^4=-\frac{\sigma K}{a^{1/2}}+\alpha^{4},
\end{eqnarray}
where $\alpha^{i}$ are functions of coordinates. Substitution of
$U_{I}^i$ in (\ref{f633})-(\ref{f6334}) yields
\begin{eqnarray}\label{f64}
\frac{d\alpha^0}{dS}=-\frac{K^2}{2}\alpha^0\alpha^1-\frac{Kr_0^{1/2}}{2}\alpha^1\alpha^4,\\
\frac{d\alpha^1}{dS}=-\frac{\sigma K^2}{2}\alpha^0-\frac{\sigma
Kr_0^{1/2}}{2}\alpha^4
+\frac{K^2}{2}\alpha^{02}+\frac{3Kr_0^{1/2}}{2}\alpha^0\alpha^4+r_0\alpha^{42},\\
\frac{d\alpha^4}{dS}=-K\frac{3-K^2r_0}{2r_0^{3/2}}\alpha^0\alpha^1-\frac{4-K^2r_0}{2r_0}\alpha^1\alpha^4.\label{f64a}
\end{eqnarray}
Equation (\ref{f41}) gives
\begin{equation}\label{f65}
0=2\sigma\alpha^0+(1-K^2r_0)\alpha^{02}-2Kr_0^{3/2}\alpha^0\alpha^4-\alpha^{12}-r_0^2\alpha^{42}.
\end{equation}

We consider case $|\alpha^{i}|<<1$ and assume $\alpha^{4}=0$ on
the surface $\chi=\pi/2$. First and second equations of system
(\ref{f64})-(\ref{f64a}) reduce to
\begin{eqnarray}\label{f66}
\frac{d\alpha^0}{dS}=-\frac{K^2}{2}\alpha^0\alpha^1,\\
\frac{d\alpha^1}{dS}=-\frac{\sigma K^2}{2}\alpha^0,\label{f66a}
\end{eqnarray}
that gives
\begin{equation}\label{f67}
\sigma\alpha^0=\frac{\alpha^{12}}{2}+H,
\end{equation}
where $H$ is constant. Substituting this expression into Eq.
(\ref{f66a}) and choosing $H=0$ we obtain
\begin{equation}\label{f68}
\frac{d\alpha^1}{dS}=-\frac{K^2\alpha^{12}}{4}.
\end{equation}

The average spacecraft's velocity on interval 20-50 a.e. is about
$\dot{r}_p=15\pm 2$ km s$^{-1}$ (See diagram in \cite{And2}) and
for approximation $S=\tau$ corresponds to $\alpha^1=5\times
10^{-5}$. Therefore this equation turns out to be
\begin{equation}\label{f69}
\ddot{r}_p=-\frac{K^2\dot{r}_p^2}{4}
\end{equation}
that yields $K=(3.94\pm 1)\times 10^{-10}$ cm$^{-1/2}$. For this
value and made choice of $S,H$ Eq. (\ref{f67}) conforms to
(\ref{f65}) without small higher-order terms.

Additional acceleration of Pioneer 11 on distance less than 20
a.e. and its predicted magnitude are contained in \rft{rva}.

{\small\begin{table}\caption{ Distance from Sun to Pioneer 11
$r_p$, in AU, its velocity $\dot{r}_p$, in km s$^{-1}$ and
unmodeled acceleration $a_p$, in $10^{-8}$ cm s$^{-2}$ (See plans
and figure in \cite{Tur}), predicted magnitude of acceleration
$|\ddot{r}_p|$, in $10^{-8}$ cm s$^{-2}$. }\label{rva}

\begin{tabular}{llllll} \noalign{\hrule height 1.5pt}

$r_p$ & 6 & 12\\ $\dot{r}_p$ & $5.7\pm 0.6$ & $12\pm 1.5$ \\ $a_p$
& $0.7\pm 1.5$ & $6.2\pm 1.9$ \\ $|\ddot{r}_p|$ & $1.26\pm 0.75$ &
$5.6\pm 3.3$ \\ \hline

\noalign{\hrule height 1.5pt}
\end{tabular}

\end{table}}

\subsection{Additional acceleration of planets}

In this section we will test proposed model by finding additional
acceleration for planets of the solar system and comparing them
with observations data. Further we will use following denotations:
$\gamma$ is gravity constant, $M$ is the Sun's mass, $\Omega$ is
its semimajor axis, $e$ is eccentricity, $n=\sqrt{\gamma
M/\Omega^{3}}$ is unperturbed Keplerian mean motion, $P=2\pi/n$ is
orbital period and $\xi$ is eccentric anomaly.

Parameters of motion are given by equations
\begin{eqnarray}\label{f80}
r=\Omega(1-e\cos\xi),\,\,\,\,\, nt=\xi-e\sin\xi .
\end{eqnarray}
Differentiation of these relations with respect to $t$ yields
\begin{eqnarray}\label{f81}
\dot{r}=\Omega e\sin\xi\dot{\xi},\,\,\,\,\,
\dot{\xi}=\frac{n}{1-e\cos\xi}
\end{eqnarray}
and radial velocity is rewritten as
\begin{equation}\label{f82}
\dot{r}=\frac{n\Omega e\sin\xi}{1-e\cos\xi}.
\end{equation}

A mean squared radial velocity during half-period
\begin{equation}\label{f86}
\left\langle\dot{r}\right\rangle=\left(\frac{2}{P}\int\limits_{0}^{P/2}\dot{r}^2
dt\right)^{1/2}
\end{equation}
after following from (\ref{f81}) substitution
\begin{equation}\label{f87}
dt=\frac{1}{n}(1-e\cos\xi)d\xi
\end{equation}
will be
\begin{equation}\label{f88}
\left\langle\dot{r}\right\rangle=\left(\frac{4\pi
e^2\Omega^2}{P^2}\int\limits_{0}^{\pi}
\frac{\sin^2\xi}{1-e\cos\xi}d\xi\right)^{1/2}.
\end{equation}

Values $\left\langle\dot{r}\right\rangle$ for the planets,
corresponding them Pioneer-like acceleration
\begin{equation}\label{f89}
A_p=-\frac{K^2\left\langle\dot{r}\right\rangle^2}{4}
\end{equation}
and anomalous accelerations of planets $A_{obs}$, obtained from
observations are in \rft{pl}. Predicted additional radial
acceleration for Yupiter, Saturn, Uranus is within the observation
error and for asteroid Icarus it is close to upper limit of
$A_{obs}$.

{\small\begin{table}\caption{Semimajor axes $\Omega$ in AU,
eccentricities $e$, orbital periods $P$ in years, mean squared
radial velocities $\left\langle\dot{r}\right\rangle$ in $10^{4}$
cm s$^{-1}$, coordinate $r$, predicted magnitude of additional
radial accelerations $|A_p|$ in $10^{-13}$ cm s$^{-2}$ and
determined from observations anomalous radial accelerations
$A_{obs}$ in $10^{-13}$ cm s$^{-2}$  for the planets \cite{Ior}
and asteroid Icarus \cite{Sand}. }\label{pl}

\begin{tabular}{lllllll} \noalign{\hrule height 1.5pt}

 & Yupiter & Saturn & Uranus & Neptune & Pluto & Icarus \\
$\Omega$ & 5.2 & 9.5 & 19.19 & 30.06 & 39.48 & 1.077
  \\ $e$ & 0.048 & 0.056 &
0.047 & 0.008 & 0.248 & 0.826\\ $P$ & 11.86 & 29.45 & 84.07 &
163.72 & 248.02 & 1.12 \\ $\left\langle\dot{r}\right\rangle$ &
4.437 & 3.809 & 2.242 & 0.3095 & 8.386 & 190.9\\ $|A_p|$ & $7.6\pm
5$ & $5.6\pm 3.5$ & $2\pm 1.2$ & $0.037\pm 0.022$ & $27\pm 16$ &
$(14.2\pm 8.4)\times 10^{5}$ \\ $A_{obs}$  & $100\pm 700$ &
$(-1.34\pm 4.23)\times 10^{4}$ & $(0.058\pm 1.338)\times 10^{5}$ &
- & - & $<6.3\times 10^{5}$
\\ \hline

\noalign{\hrule height 1.5pt}
\end{tabular}

\end{table}}

\subsection{Model with time-like fifth coordinate}

In hyperbolic coordinates in the neighborhood of point
$(\check{\tau}_0, \check{r}_0,\pi/2,0,0)$ Eqs.
(\ref{f492})-(\ref{f492a}) correspond to system
(\ref{f449})-(\ref{f449a}) reduced to
\begin{eqnarray}\label{f93}
\frac{d^{2}\check{\tau}}{dS^{2}}-\frac{\check{K}^2}{2}\frac{d\check{\tau}}{dS}\frac{d\check{a}}{dS}
-\frac{\check{K}\check{r}_0^{1/2}}{2}\frac{d\check{a}}{dS}\frac{d\check{\chi}}{dS}=0,\\
\frac{d^{2}\check{a}}{dS^{2}}+\frac{\check{K}^2}{2}\left(\frac{d\check{\tau}}{dS}\right)^{2}
+\frac{3\check{r}_0^{1/2}}{2}\frac{d\check{\tau}}{dS}\frac{d\check{\chi}}{dS}
+\check{r}_0\left(\frac{d\check{\chi}}{dS}\right)^{2}=0,\\
\frac{d^{2}\check{\chi}}{dS^{2}}+\frac{\check{K}(3
+\check{K}^{2}\check{r}_0)}{2\check{r}_0^{3/2}}\frac{d\check{\tau}}{dS}\frac{d\check{a}}{dS}
+\frac{4+\check{K}^{2}\check{r}_0}{2\check{r}_0}\frac{d\check{a}}{dS}\frac{d\check{\chi}}{dS}=0.\label{f93a}
\end{eqnarray}
For closed to hyperbolic motion (\ref{f433}) non-vanishing
five-velocities are written in form
\begin{eqnarray}\label{f94}
\check{U}_{I}^0=\sigma+\check{\alpha}^{0},\\
\check{U}_{I}^1=\check{\alpha}^{1},\\
\check{U}_{I}^4=-\frac{\sigma
\check{K}}{\check{a}^{1/2}}+\check{\alpha}^{4},
\end{eqnarray}
where $\check{\alpha}^{i}$ are functions of coordinates.
Substitution of $\check{U}_{I}^i$ in (\ref{f93})-(\ref{f93a})
yields
\begin{eqnarray}\label{f95}
\frac{d\check{\alpha}^0}{dS}=\frac{\check{K}^2}{2}\check{\alpha}^0\check{\alpha}^1
+\frac{\check{K}\check{r}_0^{1/2}}{2}\check{\alpha}^1\check{\alpha}^4,\\
\frac{d\check{\alpha}^1}{dS}=\frac{\sigma
\check{K}^2}{2}\check{\alpha}^0 +\frac{\sigma
\check{K}\check{r}_0^{1/2}}{2}\check{\alpha}^4
-\frac{\check{K}^2}{2}\check{\alpha}^{02}
-\frac{3\check{K}\check{r}_0^{1/2}}{2}\check{\alpha}^0\check{\alpha}^4-\check{r}_0\check{\alpha}^{42},\label{f95b}\\
\frac{d\check{\alpha}^4}{dS}=-\check{K}\frac{3+\check{K}^2\check{r}_0}{2\check{r}_0^{3/2}}\check{\alpha}^0\check{\alpha}^1
-\frac{4+\check{K}^2\check{r}_0}{2\check{r}_0}\check{\alpha}^1\check{\alpha}^4.\label{f95a}
\end{eqnarray}
Equation (\ref{f414}) gives
\begin{equation}\label{f96}
0=2\sigma\check{\alpha}^0+(1+\check{K}^2\check{r}_0)\check{\alpha}^{02}
+2\check{K}\check{r}_0^{3/2}\check{\alpha}^0\check{\alpha}^4-\check{\alpha}^{12}+\check{r}_0^2\check{\alpha}^{42}.
\end{equation}

We consider case $|\check{\alpha}^{i}|<<1$ and assume
$\check{\alpha}^{4}=0$ on the surface $\check{\chi}=0$. Equations
(\ref{f95}), (\ref{f95b}) reduce to
\begin{eqnarray}\label{f97}
\frac{d\check{\alpha}^0}{dS}=\frac{\check{K}^2}{2}\check{\alpha}^0\check{\alpha}^1,\\
\frac{d\check{\alpha}^1}{dS}=\frac{\sigma
\check{K}^2}{2}\check{\alpha}^0,\label{f97a}
\end{eqnarray}
that gives
\begin{equation}\label{f98}
\sigma\check{\alpha}^0=\frac{\check{\alpha}^{12}}{2}+\check{H},
\end{equation}
where $\check{H}$ is constant. Substituting this expression into
Eq. (\ref{f97a}) we obtain
\begin{equation}\label{f99}
\frac{d\check{\alpha}^1}{dS}=\frac{\check{K}^2\check{\alpha}^{12}}{4}+\frac{\check{H}\check{K}^2}{2}.
\end{equation}
This result doesn't conform to the Pioneer effect, so far as in
accordance with this expression with increase of magnitude of
radial velocity  corresponding growth of acceleration will be
positive.

For cylindrical coordinates in case $\check{V}^3=0$ in point
$(\check{\tau}_0, \check{r}_0,\pi/2,0,0)$ from Eq. (\ref{f492b})
we obtain
\begin{equation}\label{f991}
\frac{d^{2}\check{r}}{dS^{2}}=-\frac{{\check{K}^2}}{2}\left(\frac{d\check{\tau}}{dS}\right)^{2}.
\end{equation}
With condition $|\check{V}^1|<<1$, this equation conforms to
unmodeled acceleration of Pioneer 10/11 on distance 20-50 a.e.,
but gives the same acceleration for Pioneer 11 on distance less
than 20 a.e. and for planets of the Sun system that contradicts
data of observations (Tables \ref{rva},\ref{pl}).

\section{Conclusion}

Solutions of geodesics equations is found for rotating space in 5D
with angular velocity, being inversely proportional to the square
root of radius. They describe motion of particle having rest mass
in a circle with space-like fifth coordinate in spherical
(\ref{f42}-\ref{f51}) or cylindric (\ref{f10})-(\ref{f10a}) frames
and hyperbolic motion with time-like fifth coordinate
(\ref{f433})-(\ref{f4334}), (\ref{f49})-(\ref{f499}). Time
dilation is absence for solutions of the first type (\ref{f515}),
(\ref{f443}), and in Kaluza-Klein model they corresponds to
neutrally charged particle (\ref{f56}).

Proposed toy-model of particles motion is based on idea of double
manyfold Universe. It is supported by notion that closed geodesic
of elementary particle having a rest mass corresponds to motion of
the pair particle-antiparticle in mirror worlds.

Analogy with motion in central gravity field in 4D is employed for
determination of velocity and acceleration observed in 4D sheet
for particles moving in 5D bulk. We obtain approximate solution in
the neighborhood of surface with zero fifth coordinate in
cylindric frame for geodesics (\ref{f633})-(\ref{f6334}),
(\ref{f93})-(\ref{f93a}) deviating from having constant radius.
With space-like fifth coordinate a body in 4D space-time with
appropriate radial velocity will have centripetal acceleration
(\ref{f69}) being proportional to square of radius and directed
towards 5D geodesic axis. That roughly conforms to underlying
properties of the Pioneer-effect, namely, constant additional
acceleration of apparatus towards the Sun on distance from 20 to
50 a.e., its increase from 5 to 20 a.e., observed absence of one
in motion of planets.

\end{document}